\documentclass{article}
\usepackage{spconf,amsmath,graphicx}
\usepackage{subcaption}
\usepackage{makecell}
\usepackage[skip=3pt]{caption} 
\usepackage{multirow}
\usepackage{float}
\usepackage{subcaption}
\usepackage{tipa}
\usepackage{cite}
\usepackage{amsmath,graphicx}
\usepackage[table,xcdraw]{xcolor}
\usepackage{bm}
\usepackage{listings}
\DeclareMathOperator*{\argmax}{argmax}

\title{Data augmentation with Locally-time Reversed Speech \\for automatic speech Recognition}

\name{Si-Ioi Ng, Tan Lee}
\address{
  Department of Electronic Engineering \\ The Chinese University of Hong Kong}

\begin{document}
\maketitle

\begin{abstract}
Psychoacoustic studies have shown that locally-time reversed (LTR) speech, i.e., signal samples time-reversed within a short segment, can be accurately recognised by human listeners. This study addresses the question of how well a state-of-the-art automatic speech recognition (ASR) system would perform on LTR speech. The underlying objective is to explore the feasibility of deploying LTR speech in the training of end-to-end (E2E) ASR models, as an attempt to data augmentation for improving the recognition performance. The investigation starts with experiments to understand the effect of LTR speech on general-purpose ASR. LTR speech with reversed segment duration of 5 ms - 50 ms is rendered and evaluated. For ASR training data augmentation with LTR speech, training sets are created by combining natural speech with different partitions of LTR speech. The efficacy of data augmentation is confirmed by ASR results on speech corpora in various languages and speaking styles. ASR on LTR speech with reversed segment duration of 15 ms - 30 ms is found to have lower error rate than with other segment duration. Data augmentation with these LTR speech achieves satisfactory and consistent improvement on ASR performance.



\end{abstract}



\section{Introduction}\label{intro} 
Perception and recognition of speech requires tremendous amount of processing in human brain.
Cross-disciplinary studies investigated the relation between speech sound and human behaviour. 
To understand how human listeners response to different speech stimuli, unintelligible speech is often used.
Speech reversal refers to the operation of reversing a speech signal, which renders an audio signal that may contain unintelligible speech.
In studies related to irrelevant speech effect (IFS), \cite{ellermeier2014psychoacoustics}, 
time-reversed speech was found to affect short-term memory of human listeners.
In neuro-physiological research \cite{howard2010discrimination}, when subjects listened to time-reversed speech, observation on neural responses revealed that human brain discriminates speech stimuli based on acoustics instead of linguistic features.
In relation to speech technology, time-reversed speech was applied to train speech enhancement model \cite{chao2021tenet} to improve the quality of enhanced speech.

While time-reversed speech is generally unintelligible, speech intelligibility, referring to percentage of correctly recognised syllables or words, was found to withstand time reversal performed in local segments \cite{saberi1999cognitive}. This is known as locally-time reversed (LTR) speech. Let a 5-second speech utterance be divided into 250 segments of 20 ms in length. If the samples in each segment are reversed in time order, listeners still consider the manipulated speech highly intelligible. Cognitive restoration of LTR speech was further investigated in a multilingual study \cite{ueda2017intelligibility}. 
The speech intelligibility degraded with the duration of time-reversed segment increasing across different languages.
The intelligibility remains at above 90\% as the reversed segment duration is shorter than 50 ms, and drops to below 25\% as the duration is longer than 100 ms. This observation was in agreement with other monolingual studies on LTR speech \cite{greenberg2001relation, ishida2018perceptual, magrin2002intelligibility, kiss2008auditory}. 

Being analogous to human speech recognition (HSR), ASR aims to transform human speech into text. It could be cast as a task of pattern recognition.
Motivated by human perception of LTR speech, we are interested in asking the following questions:
how would a general-purpose ASR system perform on LTR speech? 
Apparently, local time reversal alters the spectral characteristics of natural speech. 
The alteration increases the discrepancy between training data of natural speech and test data of LTR speech. 
If reversed segment duration is relatively short, e.g. below 50 ms, acoustical characteristics of phonemes, syllable structure, word order, 
are largely preserved in LTR speech. 
We expect the natural and the LTR speech are similar. The ASR performance on LTR speech would be close to that on natural speech. 

State-of-the-art ASR systems are predominately based on the data-driven deep neural network (DNN) approach. 
Using data augmentation strategies to escalate the quantity and variability of training data is effective to improve the performance of DNN models. Data augmentation does not require additional acoustic data, and is fairly easy to be implemented. 
Commonly used methods include vocal-tract length perturbation \cite{jaitly2013vocal}, speed perturbation \cite{ko2015audio}, spectral masking, swapping and stretching \cite{Park2019, song2020specswap, nguyen2020improving}. 
Data augmentation with LTR speech has the potential to help improve ASR performance.


The present study starts with an investigation on the effect of LTR speech on ASR.  
The performance of state-of-the-art end-to-end (E2E) ASR systems on LTR speech is evaluated with speech corpora of different languages and speaking styles. 
The E2E design is chosen as it was shown to better match human performance than other predominant systems in psychometric experiments \cite{weerts2021psychometrics}.
We find 
recognition on LTR speech with particularly short and particularly long reversed segment duration is prone to erroneous results.
Based on the observation on ASR for LTR speech, 
we propose a speech data augmentation method, which
combines LTR speech with natural speech in ASR training.
Extensive ASR experiments are carried out to confirm the effectiveness of this strategy.

\section{Method}
\vspace{-2mm}
\subsection{Speech corpora}

\begin{table}[t!]
\caption{Speech corpora.}
\centering
\resizebox{0.98\linewidth}{!}{
\begin{tabular}{|c|c|c|c|}
\hline
\textbf{Corpus} & \textbf{Type} & \textbf{Language} & \textbf{Hours} \\ \hline 
TIMIT     & Read        & English          & 5.4 \\ \hline  
WSJ       & Read        & English          & 81  \\ \hline 
Aishell   & Read        & Mandarin Chinese & 178 \\ \hline  
Ted-lium-2 & Spontaneous & English          & 118 \\ \hline 
\end{tabular}
}
\label{dataset}
\end{table}

ASR experiments are carried out on natural and LTR speech with four speech databases. They are TIMIT \cite{garofolo1993darpa}, WSJ \cite{paul1992design}, Aishell \cite{bu2017aishell}, Ted-lium2 \cite{rousseau2012ted}. The databases are chosen based on the consideration of language diversity, speaking style and amount of speech data.
TIMIT is a read-speech corpus of English. The corpus contains phonetically-balanced utterances. It was widely used to support research on acoustic-phonetics and continuous speech recognition.
WSJ is a read-speech corpus of English news, covering a large vocabulary size. The corpus was extensively used for benchmark evaluation of ASR systems.
Aishell is an open-sourced corpus of Mandarin read-speech. 
The content covers a wide range of topics. 
Ted-lium2 is a corpus of spontaneous English speech. The speech was extracted from recordings of public talks, and the content was not well organized. 
Details of the four databases are summarised as in Table \ref{dataset}. 

\vspace{-2mm}
\subsection{Locally-time reversed speech}


\begin{figure}[t!]
        \setlength\belowcaptionskip{-0.25\baselineskip}
        \centering
        \begin{subfigure}[b]{0.38\textwidth}
        \includegraphics[width=\textwidth]{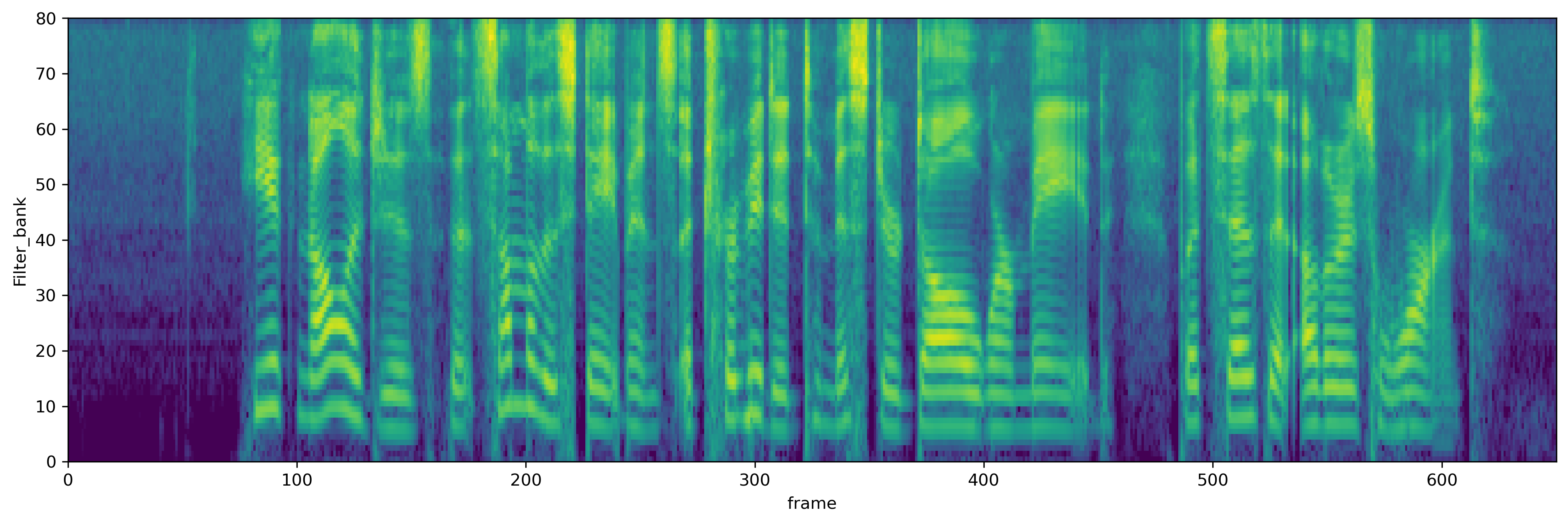}
        \caption{Unaltered utterance}
        \end{subfigure}

        \begin{subfigure}[b]{0.38\textwidth}
        \includegraphics[width=\textwidth]{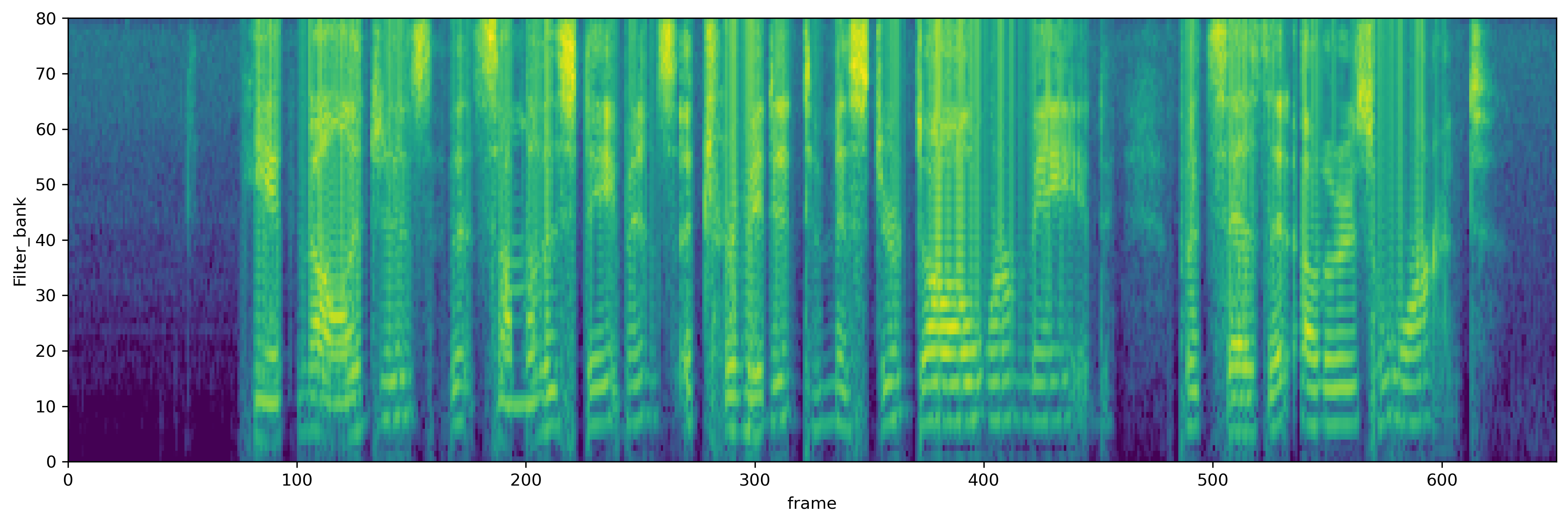}
        \caption{Reversed segment duration: 5ms.}
        \end{subfigure}
        
        
        \begin{subfigure}[b]{0.38\textwidth}
        \includegraphics[width=\textwidth]{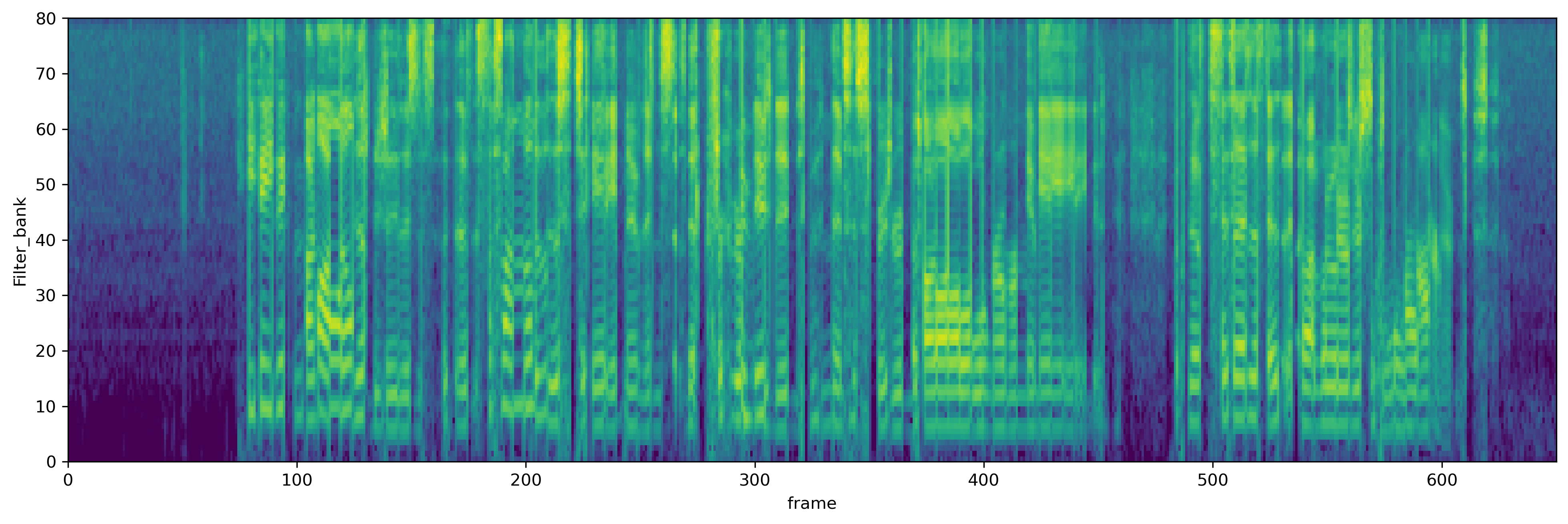}
        \caption{Reversed segment duration: 50 ms.}
        \end{subfigure}

        \caption{Spectrogram of an utterance from WSJ.}
        \label{spectro}
\end{figure}

The narrowing performance gap between ASR and HSR has motivated exploratory studies on the proximity of ASR and human auditory system \cite{lippmann1997speech, thomas2019english}.
Despite that some of the latest ASR systems outperform human performance on specific tasks \cite{xiong2017toward,nguyen2020improving}, ASR performance is clearly inferior to human auditory system under challenging acoustic conditions, i.e. when input speech is clipped, spectrally modulated, band-pass filtered, or masked by noise \cite{weerts2021psychometrics,spille2018comparing}. 
The present study is focused on the effect of LTR speech on E2E ASR. LTR speech is a type of temporally distorted speech. Local reversal with segment duration from 5 ms to 50 ms, in the step size of 5 ms, is applied to the speech data in TIMIT, WSJ, Aishell and Ted-lium2.
In \cite{ashihara21_interspeech}, the effect of LTR speech on ASR
was investigated with the English corpus Ted-lium2 and the Japanese corpus CSJ \cite{maekawa2000spontaneous}. 
LTR speech with segment duration of 25 ms, 50 ms, 70 ms and 100 ms were attempted.
The ASR performance was found to degrade substantially when the reversed segment duration is longer than 50 ms.

Figure \ref{spectro} illustrates the spectrograms of natural and LTR speech of the same utterance.
It is noted that the temporal order of phonemes is mostly preserved in the LTR speech. 
In the case of 5 ms local reversal, the acoustic properties of speech, e.g. pitch and formant contours, are locally altered.
Local reversal in short segments tends to cause more noise in speech. 
This is related to the discontinuities of speech samples at the segment boundaries. 
Reversal in relatively longer segments, e.g. 50 ms, does not cause much noise, but
inclines to disrupt the smooth transition between speech sounds. 

\subsection{E2E ASR system}
The ASR model is implemented with an encoder-decoder architecture. It is trained toward two objectives, namely the connectionist temporal classification (CTC) and the attention-based classification \cite{kim2017joint}. 
Tokens such as English/Chinese characters, phonemes and graphemes, are extracted from given speech transcriptions as the output labels (or symbols).
CTC aims to map input acoustic features to output token sequence via the encoder. 
Assuming a ground-truth token sequence $\mathbf{y}$ has $K$ distinct token labels, an intermediate token sequence $\pi=(\pi_1, ..., \pi_T)$ of length $T$ is used to represent $\mathbf{y}$ with repetitions of token labels and blank label, where $\pi_t\in\{1,...,K\}\cup\{Blank\}$.
Given input features $\mathbf{x}$ of $T$ frames in length, 
the CTC loss function is defined as,
\begin{equation*}
     \mathcal{L}_{CTC}= 
     -\log{{P}_{CTC}(\mathbf{y}|\mathbf{x})}, = -\log{\sum_{\mathbf{\pi}\in \Phi{(\mathbf{y'}})} P(\mathbf{\pi}|\mathbf{x})},
\end{equation*}
where $\mathbf{y'}$ is the prolonged version of $\mathbf{y}$ with blank tokens inserted in-between ground-truth tokens, and at the beginning and the end of $\mathbf{y}$. $\Phi{(\mathbf{y'})}$ denotes all possible sequences of $\mathbf{y'}$.

The decoder predicts output token sequence in an auto-regressive manner with the attention mechanism. Let $y_u$ be the u-th ground-truth token, the decoder loss is expressed as,
\begin{small}
\begin{equation*}
    \mathcal{L}_{att} = -\log{{P}_{att}(\mathbf{y}|\mathbf{x})} = -\sum_{u}\log{P(y_u|\mathbf{x},y_{1:u-1})},
\end{equation*}
\end{small}
The two loss functions are fused in a multi-task manner by,
\begin{equation*}
     \mathcal{L}_{MTL} = \lambda \mathcal{L}_{CTC} + (1-\lambda) \mathcal{L}_{Att},
\end{equation*}
where $\lambda \in [0,1]$ is the task weight.
In the decoder, the language model (LM) can be incorporated by shallow fusion. 
The ASR system aims to determine the best token sequence $\mathbf{\hat{y}}$ as,
\begin{multline*}
\mathbf{\hat{y}} = \argmax_{\mathbf{y}\in Y^*}\{ \alpha\log{{P}_{CTC}(\mathbf{y}|\mathbf{x})} + (1-\alpha)\log{{P}_{att}(\mathbf{y}|\mathbf{x})} \\ + \beta\log{{P}_{LM}(\mathbf{y})}\},
\end{multline*}
where $Y^*$ denote the complete set of hypotheses.
$\alpha$ and $\beta$ are hyper-parameters that control the contributions of CTC, attention and language model (LM). 
\begin{table}[t!]
\caption{System and training configurations of E2E ASR for the four tasks/databases.}
\centering
\resizebox{0.99\linewidth}{!}{
\begin{tabular}{|c|c|c|c|c|}
\hline
\textbf{Corpus} & \textbf{Model}                                                               & \textbf{\begin{tabular}[c]{@{}c@{}}Architecture \\ (Layer / Units)\end{tabular}} & \textbf{\begin{tabular}[c]{@{}c@{}}CTC-\\ Attention\\ Weight\end{tabular}} & \textbf{\begin{tabular}[c]{@{}c@{}}Epochs /\\ Patience\end{tabular}} \\ \hline
TIMIT           & \begin{tabular}[c]{@{}c@{}}Bi-directional GRU\\ with projection\end{tabular} & \begin{tabular}[c]{@{}c@{}}Encoder: 5 / 320\\ Decoder: 1  / 300 \end{tabular}                                                                          & 0.5                                                                        & 20 / 3                                                               \\ \hline
WSJ             & Conformer \cite{gulati20_interspeech}                                                                   & \begin{tabular}[c]{@{}c@{}}Encoder: 12 / 2048\\ Decoder: 6 / 2048\end{tabular}   & 0.3                                                                        & 100 / 5                                                              \\ \hline
Aishell         & Transformer                                                                  & \begin{tabular}[c]{@{}c@{}}Encoder: 12 / 2048\\ Decoder: 6  / 2048\end{tabular}  & 0.3                                                                        & 100 / None                                                           \\ \hline
Ted-lium2       & Conformer                                                                    & \begin{tabular}[c]{@{}c@{}}Encoder: 17 / 1024 \\ Decoder: 4 / 1024\end{tabular}  & 0.3                                                                        & 50 / None                                                            \\ \hline
\end{tabular}
}
\label{models}
\end{table}

\begin{table}[t!]
\caption{RNNLM training setup.}
\centering
\resizebox{0.90\linewidth}{!}{
\begin{tabular}{|c|c|c|c|}
\hline
\textbf{Corpus} & \textbf{Model}                                                                              & \textbf{\begin{tabular}[c]{@{}c@{}}Architecture \\ (Layer / Units)\end{tabular}} & \textbf{\begin{tabular}[c]{@{}c@{}}Epochs /\\ Patience\end{tabular}} \\ \hline
WSJ             & \multirow{3}{*}{\begin{tabular}[c]{@{}c@{}}Long-short term memory\\ (LSTM)\end{tabular}} & 1 / 1000                                                                         & 20 / 3                                                               \\ \cline{1-1} \cline{3-4} 
Aishell         &                                                                                             & 2 / 650                                                                          & 20 / 3                                                               \\ \cline{1-1} \cline{3-4} 
Ted-lium2       &                                                                                             & 4 / 2048                                                                         & 2 / None                                                             \\ \hline
\end{tabular}
}
\label{RNNLM}
\end{table}

Training of the E2E models is implemented with the ESPNET toolkit \cite{watanabe18_interspeech} with the Adadelta optimizer \cite{zeiler2012adadelta}. The input acoustic features are 80-dimensional Mel-scale filter-bank (F-bank) coefficients with mean and variance normalised.
The model architectures and training parameters are summarised as in Table \ref{models}.
Recurrent neural network LMs (RNNLMs) are trained on WSJ, Aishell, and Ted-lium2. All RNNLMs adopt the architecture of long-short term memory (LSTM). Details are provided as in Table \ref{RNNLM}. 
In addition, a 4-gram LM is trained on Aishell and jointly used with the RNNLM. 

\vspace{-1mm}
\section{Results And Discussion}
\vspace{-1mm}
\subsection{ASR performance on locally time-reversed speech}
\begin{figure}[t!]
        \setlength\belowcaptionskip{-0.25\baselineskip}
       \centering
\begin{subfigure}[b]{0.235\textwidth}
\includegraphics[width=\textwidth]{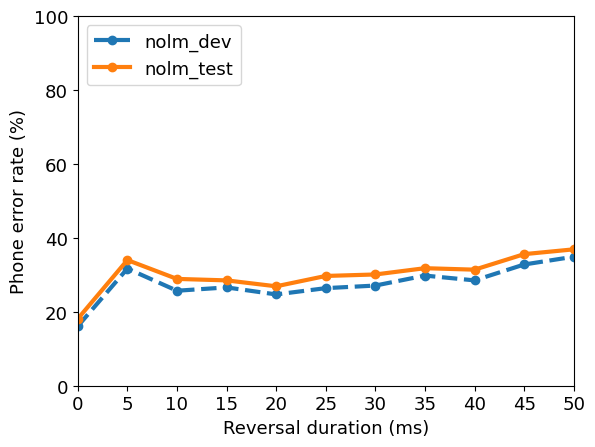}
\caption{TIMIT}
\end{subfigure}
\begin{subfigure}[b]{0.235\textwidth}
\includegraphics[width=\textwidth]{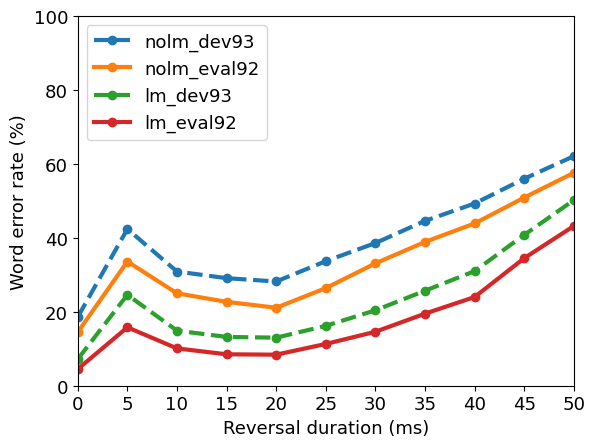}
\caption{WSJ}
\end{subfigure}
\begin{subfigure}[b]{0.235\textwidth}
\includegraphics[width=\textwidth]{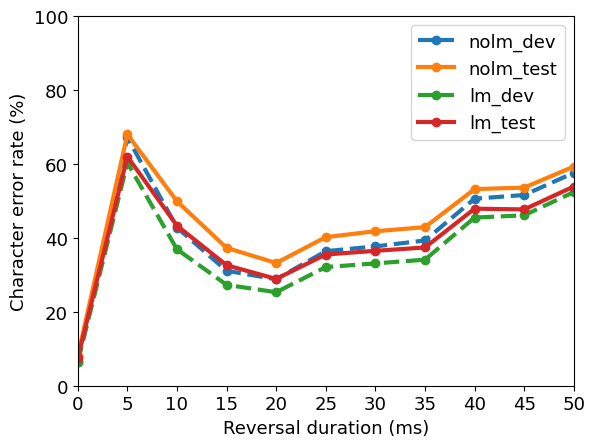}
\caption{Aishell}
\end{subfigure}
\begin{subfigure}[b]{0.24\textwidth}
\includegraphics[width=\textwidth]{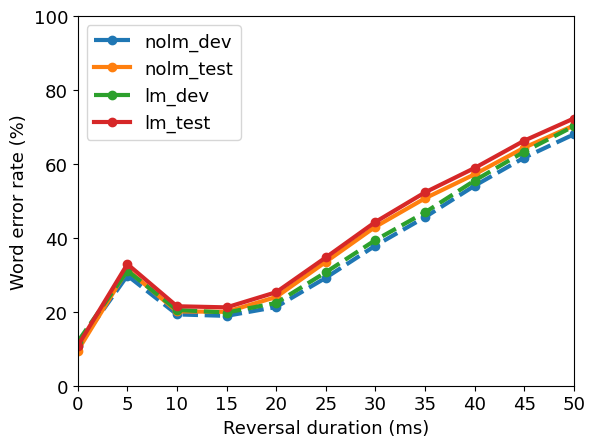}
\caption{Ted-lium2}
\end{subfigure}
\caption{ASR performance on LTR speech in different languages}
\label{reverse results}
\end{figure}



The ASR systems are first trained on natural speech and used to decode input utterances of LTR speech. The segment duration for local reversal varies from 5 ms to 50 ms. 
The recognition performance is measured in terms of phone error rate (PER) in TIMIT, character error rate (CER) in Aishell, and word error rate (WER) in WSJ and Ted-lium2.
Except the TIMIT phone recognition task, shallow fusion with RNNLM is applied in the decoding. The results are shown as in Figure \ref{reverse results}.
It must be noted that many previous studies on LTR speech \cite{ueda2017intelligibility, greenberg2001relation, ishida2018perceptual,magrin2002intelligibility,ashihara21_interspeech} assumed minimum segment duration of 20 ms. The effect of local time reversal in notably short segments, i.e. below 20 ms, has not been well understood. 

\begin{table*}[hbt!]
\centering
\caption{Recognition performance using LTR speech in ASR training.}
\resizebox{\linewidth}{!}{
\begin{tabular}{|c|c|c|c|ccccc|cc|}
\hline
\multirow{2}{*}{\textbf{Corpus}}                                                  & \multirow{2}{*}{\textbf{Metric}} & \multirow{2}{*}{\textbf{Corpus}} & \multicolumn{8}{c|}{\textbf{Training set}}                                                                                                                                                                                                                                                                                                                                                                                                                                                                                                                    \\ \cline{4-11} 
                                                                                  &                                  &                                  & \begin{tabular}[c]{@{}c@{}}Original\\ (Baseline)\end{tabular} & \begin{tabular}[c]{@{}c@{}}Original\\  + 5ms + 10ms\\ (Set 1)\end{tabular} & \begin{tabular}[c]{@{}c@{}}Original\\  + 15ms + 20ms\\ (Set 2)\end{tabular} & \begin{tabular}[c]{@{}c@{}}Original\\  + 25ms + 30ms \\ (Set 3)\end{tabular} & \begin{tabular}[c]{@{}c@{}}Original\\  + 35ms + 40ms \\ (Set 4)\end{tabular} & \begin{tabular}[c]{@{}c@{}}Original\\  + 45ms +50ms \\ (Set 5)\end{tabular} & \begin{tabular}[c]{@{}c@{}}Speed \\ Perterbation\\ (3-fold)\end{tabular} & SpecAugment \\ \hline
\begin{tabular}[c]{@{}c@{}}Timit\\ (dev / test)\end{tabular}                      & PER(\%)                          & No                               & 16.1 / 18.1                                                   & 16.1 / 18.3                                                                & 15.8 / 17.6                                                                 & 16.1 / 17.9                                                                  & 16.5 / 18.3                                                                  & 17.4 / 19.1                                                                 & 14.9 / 17.3                                                              & 17.8 / 18.8 \\ \hline
\multirow{2}{*}{\begin{tabular}[c]{@{}c@{}}WSJ\\ (dev93 / eval92)\end{tabular}}   & \multirow{2}{*}{WER(\%)}         & Yes                              & 7.2 / 4.5                                                     & 7.0 / 4.6                                                                  & 7.0 / 4.5                                                                   & 6.4 / 4.3                                                                    & 6.8 / 4.4                                                                    & 6.9 / 4.4                                                                   & 6.1 / 3.9                                                                & 7.4 / 4.3   \\
                                                                                  &                                  & No                               & 18.6 / 14.5                                                   & 13.2 / 9.9                                                                 & 14.0 / 11.1                                                                 & 13.6 / 10.5                                                                  & 14.1 / 10.3                                                                  & 13.8 / 10.7                                                                 & 12.2 / 9.7                                                               & 15.3 / 11.9 \\ \hline
\multirow{2}{*}{\begin{tabular}[c]{@{}c@{}}Aishell\\ (dev / test)\end{tabular}}   & \multirow{2}{*}{CER(\%)}         & Yes                              & 6.4 / 7.4                                                     & 5.7 / 6.3                                                                  & 5.7 / 6.5                                                                   & 5.8 / 6.5                                                                    & 5.8 / 6.7                                                                    & 5.9 / 6.7                                                                   & 5.4 / 5.9                                                                & 6.5 / 7.0   \\
                                                                                  &                                  & No                               & 6.5 / 8.0                                                     & 5.8 / 6.6                                                                  & 6.5 / 6.8                                                                   & 5.9 / 6.9                                                                    & 6.0 / 7.1                                                                    & 6.1 / 7.1                                                                   & 5.7 / 6.7                                                                & 5.1 / 5.8   \\ \hline
\multirow{2}{*}{\begin{tabular}[c]{@{}c@{}}Ted-lium2\\ (dev / test)\end{tabular}} & \multirow{2}{*}{WER(\%)}         & Yes                              & 10.4 / 8.4                                                    & 10.0 / 8.3                                                                 & 10.2 / 8.3                                                                  & 10.2 / 8.1                                                                   & 10.2 / 8.1                                                                   & 10.1 / 8.2                                                                  & 9.2 / 7.7                                                                & 9.8 / 7.5   \\
                                                                                  &                                  & No                               & 10.8 / 9.5                                                    & 10.2 / 9.1                                                                 & 10.3 / 9.0                                                                  & 11.0 / 9.0                                                                   & 10.9 / 9.0                                                                   & 10.4 / 9.0                                                                  & 9.8 / 8.3                                                                & 10.6 / 8.0  \\ \hline
\end{tabular}
}
\label{Final performance}
\end{table*}

The error rate increases first from no reversal to reversal with 5 ms segment duration. When the segment duration further increases to 10 ms and 20 ms, the error rate drops. When the segment duration increases from 20 to 50 ms, the error rate goes up steadily. These trends are consistent among different corpora regardless of the model architecture. 
The error rate peak at 5 ms could be caused by the additional noise and changes of acoustic properties in LTR speech. 
The dissimilarity between training and test data deteriorates the recognition performance. 





Different types of ASR errors are analysed for LTR speech and natural speech.
In the case of TIMIT, 
there are frequent errors concerning nasal substitution (/m/ \textrightarrow{} /n/), and approximant substitution (/w/ \textrightarrow{} /l/). These errors are not common in natural speech.
Insertion and deletion of the article `a' in WSJ, and deletion of `a', `and', 'the' in Ted-lium2, are persistent ASR errors for natural speech. 
These errors are found more frequently for LTR speech, especially when the reversed segment duration is 5 ms or above 20 ms.
For Aishell, substitution errors by homophones are common in the recognition of natural speech, i.e., two different Chinese characters share the same constituent phonemes and tone. 
For LTR speech, ASR is more prone to minimal-pair errors when the reversed segment duration is 10 ms or below. Two Chinese characters are said to be a minimal pair if they differ by a single phonological element. For instance, the two differ in tone, e.g. `shi4' (city) and `shi2' (time); or differ in one phoneme, e.g. Mandarin particles of `le' and `de'. 
More deletion errors are found in ASR when the reversed segment duration is 15 ms or above. 

The recognition performance and error analysis of LTR speech
help understand the relation between LTR speech and general-purpose ASR systems.
The results suggest that if appropriate reversed segment duration is chosen, the mismatch between natural and LTR speech can be alleviated. 

\subsection{Data augmentation with LTR speech for ASR training}

Utterances of LTR speech are divided into 5 sets according to the reversed segment duration, as shown in Table \ref{Final performance}.
Each set represents different levels of intelligibility. 
As an attempt to data augmentation, each set of LTR speech utterances are combined with natural speech in ASR training, creating a 3-fold training set. 
Data augmentation with LTR speech is compared with speed perturbation and SpecAugment, which are the most widely used strategies of data augmentation.
With speed perturbation factor of 0.9, 1.0 and 1.1, 3-fold training set is created, and has the same quantity of training data as data augmentation with LTR speech.
SpecAugment warps and masks the time and frequency domain of the input F-bank coefficients. It does not increase the size of original training data. 
The model training follows the configurations in Table \ref{models}. 
The systems trained on natural speech are regarded as the baseline systems.
Recognition performance on natural speech are given as in Table \ref{Final performance}.

Data augmentation with LTR speech improves the ASR performance over the baseline system on all the four corpora. 
This confirm the feasibility and efficacy of deploying LTR speech in ASR training. 
For each corpus, the best performance attained by data augmentation with LTR speech is comparable to data augmentation with speed perturbation.
Data augmentation with LTR speech outperforms SpecAugment on TIMIT, WSJ and Aishell, but not on Ted-lium2.  

When shallow fusion with LM is applied, 
using LTR speech with reversed segment duration shorter than 30 ms (Set 1-3) performs better than with longer segment duration (Set 4-5) on WSJ, Aishell and Ted-lium2. Without shallow fusion, more satisfactory performance is obtained via LTR speech with reversed segment duration of 5 ms and 10 ms (Set 1). For TIMIT, the best result is achieved by using LTR speech reversed segment duration of 15 and 20 ms (Set 2). 
Consistent improvement is observed in using LTR speech with reversed segment duration of 15 ms - 30 ms (Set 2-3).
These results are in agreement with Figure \ref{reverse results}, in which the error rates are comparatively low.
LTR speech with this range of reversal contains less noise, and retains smoother transition of speech sounds than longer segment duration. Local reversal with 15 ms - 30 ms is recommended for efficacious data augmentation with LTR speech.

\vspace{-1mm}
\section{Conclusion}
\vspace{-1mm}
The effect of LTR speech on E2E ASR performance, and the feasibility of using LTR speech in ASR training are studied in this paper. 
Using general-purpose ASR systems trained on natural speech, recognition on LTR speech with relatively short and long reversed segment duration is prone to erroneous results. 
Data augmentation with LTR speech is confirmed to improve ASR performance on natural speech. 
Using LTR speech with reversed segment duration of 15 ms - 30 ms consistently improves ASR performance on corpora of different languages and speaking styles.
In future work, the intelligibility of LTR speech with short reversed segment duration, e.g. below 20 ms, will be studied via subjective rating. The use of LTR speech will be generalized to other topics in speech technology, such as speaker verification, speech enhancement and self-supervised learning of speech representation.

\small
\bibliographystyle{IEEEtran}

\bibliography{mybib}

\begin{thebibliography}{10}
\providecommand{\url}[1]{#1}
\csname url@samestyle\endcsname
\providecommand{\newblock}{\relax}
\providecommand{\bibinfo}[2]{#2}
\providecommand{\BIBentrySTDinterwordspacing}{\spaceskip=0pt\relax}
\providecommand{\BIBentryALTinterwordstretchfactor}{4}
\providecommand{\BIBentryALTinterwordspacing}{\spaceskip=\fontdimen2\font plus
\BIBentryALTinterwordstretchfactor\fontdimen3\font minus
  \fontdimen4\font\relax}
\providecommand{\BIBforeignlanguage}[2]{{%
\expandafter\ifx\csname l@#1\endcsname\relax
\typeout{** WARNING: IEEEtran.bst: No hyphenation pattern has been}%
\typeout{** loaded for the language `#1'. Using the pattern for}%
\typeout{** the default language instead.}%
\else
\language=\csname l@#1\endcsname
\fi
#2}}
\providecommand{\BIBdecl}{\relax}
\BIBdecl

\bibitem{ellermeier2014psychoacoustics}
W.~Ellermeier and K.~Zimmer, ``The psychoacoustics of the irrelevant sound
  effect,'' \emph{Acoustical Science and Technology}, vol.~35, no.~1, pp.
  10--16, 2014.

\bibitem{howard2010discrimination}
M.~F. Howard and D.~Poeppel, ``Discrimination of speech stimuli based on
  neuronal response phase patterns depends on acoustics but not
  comprehension,'' \emph{Journal of neurophysiology}, vol. 104, no.~5, pp.
  2500--2511, 2010.

\bibitem{chao2021tenet}
F.-A. Chao, S.-W.~F. Jiang, B.-C. Yan, J.-w. Hung, and B.~Chen, ``Tenet: A
  time-reversal enhancement network for noise-robust asr,'' \emph{arXiv
  preprint arXiv:2107.01531}, 2021.

\bibitem{saberi1999cognitive}
K.~Saberi and D.~R. Perrott, ``Cognitive restoration of reversed speech,''
  \emph{Nature}, vol. 398, no. 6730, pp. 760--760, 1999.

\bibitem{ueda2017intelligibility}
K.~Ueda, Y.~Nakajima, W.~Ellermeier, and F.~Kattner, ``Intelligibility of
  locally time-reversed speech: A multilingual comparison,'' \emph{Scientific
  reports}, vol.~7, no.~1, pp. 1--8, 2017.

\bibitem{greenberg2001relation}
S.~Greenberg and T.~Arai, ``The relation between speech intelligibility and the
  complex modulation spectrum,'' in \emph{Proc. of Eurospeech}, 2001.

\bibitem{ishida2018perceptual}
M.~Ishida, T.~Arai, and M.~Kashino, ``Perceptual restoration of temporally
  distorted speech in l1 vs. l2: Local time reversal and modulation
  filtering,'' \emph{Frontiers in psychology}, vol.~9, p. 1749, 2018.

\bibitem{magrin2002intelligibility}
I.~Magrin-Chagnolleau, M.~Barkat, and F.~Meunier, ``Intelligibility of reverse
  speech in french: a perceptual study,'' in \emph{Proc. of ICSLP}, 2002.

\bibitem{kiss2008auditory}
M.~Kiss, T.~Cristescu, M.~Fink, and M.~Wittmann, ``Auditory language
  comprehension of temporally reversed speech signals in native and non-native
  speakers,'' \emph{Acta neurobiologiae experimentalis}, vol.~68, no.~2, p.
  204, 2008.

\bibitem{jaitly2013vocal}
N.~Jaitly and G.~E. Hinton, ``Vocal tract length perturbation (vtlp) improves
  speech recognition,'' in \emph{Proc. of ICML}, vol. 117, 2013, p.~21.

\bibitem{ko2015audio}
T.~Ko, V.~Peddinti, D.~Povey, and S.~Khudanpur, ``Audio augmentation for speech
  recognition,'' in \emph{Proc. of Interspeech}, 2015, pp. 3586--3589.

\bibitem{Park2019}
D.~S. Park, W.~Chan, Y.~Zhang, C.-C. Chiu, B.~Zoph, E.~D. Cubuk, and Q.~V. Le,
  ``{SpecAugment: A Simple Data Augmentation Method for Automatic Speech
  Recognition},'' in \emph{Proc. of Interspeech}, 2019, pp. 2613--2617.

\bibitem{song2020specswap}
X.~Song, Z.~Wu, Y.~Huang, D.~Su, and H.~Meng, ``Specswap: A simple data
  augmentation method for end-to-end speech recognition.'' in \emph{Proc. of
  Interspeech}, 2020, pp. 581--585.

\bibitem{nguyen2020improving}
T.-S. Nguyen, S.~Stueker, J.~Niehues, and A.~Waibel, ``Improving
  sequence-to-sequence speech recognition training with on-the-fly data
  augmentation,'' in \emph{Proc. of ICASSP}, 2020, pp. 7689--7693.

\bibitem{weerts2021psychometrics}
L.~Weerts, S.~Rosen, C.~Clopath, and D.~F. Goodman, ``The psychometrics of
  automatic speech recognition,'' \emph{bioRxiv}, 2021.

\bibitem{garofolo1993darpa}
J.~S. Garofolo, L.~F. Lamel, W.~M. Fisher, J.~G. Fiscus, and D.~S. Pallett,
  ``Darpa timit acoustic-phonetic continous speech corpus cd-rom. nist speech
  disc 1-1.1,'' \emph{NASA STI/Recon technical report n}, vol.~93, p. 27403,
  1993.

\bibitem{paul1992design}
D.~B. Paul and J.~Baker, ``The design for the wall street journal-based csr
  corpus,'' in \emph{Proceedings of the Workshop on Speech and Natural
  Language}, 1992, pp. 357--362.

\bibitem{bu2017aishell}
H.~Bu, J.~Du, X.~Na, B.~Wu, and H.~Zheng, ``Aishell-1: An open-source mandarin
  speech corpus and a speech recognition baseline,'' in \emph{Proc. of
  O-COCOSDA}, 2017, pp. 1--5.

\bibitem{rousseau2012ted}
A.~Rousseau, P.~Del{\'e}glise, and Y.~Esteve, ``Ted-lium: an automatic speech
  recognition dedicated corpus.'' in \emph{LREC}, 2012, pp. 125--129.

\bibitem{lippmann1997speech}
R.~P. Lippmann, ``Speech recognition by machines and humans,'' \emph{Speech
  communication}, vol.~22, no.~1, pp. 1--15, 1997.

\bibitem{thomas2019english}
S.~Thomas, M.~Suzuki, Y.~Huang, G.~Kurata, Z.~Tuske, G.~Saon, B.~Kingsbury,
  M.~Picheny, T.~Dibert, A.~Kaiser-Schatzlein \emph{et~al.}, ``English
  broadcast news speech recognition by humans and machines,'' in \emph{Proc. of
  ICASSP}, 2019, pp. 6455--6459.

\bibitem{xiong2017toward}
W.~Xiong, J.~Droppo, X.~Huang, F.~Seide, M.~L. Seltzer, A.~Stolcke, D.~Yu, and
  G.~Zweig, ``Toward human parity in conversational speech recognition,''
  \emph{IEEE/ACM Transactions on Audio, Speech, and Language Processing},
  vol.~25, no.~12, pp. 2410--2423, 2017.

\bibitem{spille2018comparing}
C.~Spille, B.~Kollmeier, and B.~T. Meyer, ``Comparing human and automatic
  speech recognition in simple and complex acoustic scenes,'' \emph{Computer
  Speech \& Language}, vol.~52, pp. 123--140, 2018.

\bibitem{ashihara21_interspeech}
T.~Ashihara, T.~Moriya, and M.~Kashino, ``Investigating the impact of spectral
  and temporal degradation on end-to-end automatic speech recognition
  performance,'' in \emph{Proc. of Interspeech}, 2021, pp. 1757--1761.

\bibitem{maekawa2000spontaneous}
K.~Maekawa, H.~Koiso, S.~Furui, and H.~Isahara, ``Spontaneous speech corpus of
  japanese.'' in \emph{LREC}, vol.~6.\hskip 1em plus 0.5em minus 0.4em\relax
  Citeseer, 2000, pp. 1--5.

\bibitem{kim2017joint}
S.~Kim, T.~Hori, and S.~Watanabe, ``Joint ctc-attention based end-to-end speech
  recognition using multi-task learning,'' in \emph{Proc. of ICASSP}, 2017, pp.
  4835--4839.

\bibitem{gulati20_interspeech}
A.~Gulati, J.~Qin, C.-C. Chiu, N.~Parmar, Y.~Zhang, J.~Yu, W.~Han, S.~Wang,
  Z.~Zhang, Y.~Wu, and R.~Pang, ``{Conformer: Convolution-augmented Transformer
  for Speech Recognition},'' in \emph{Proc. of Interspeech}, 2020, pp.
  5036--5040.

\bibitem{watanabe18_interspeech}
S.~Watanabe, T.~Hori, S.~Karita, T.~Hayashi, J.~Nishitoba, Y.~Unno, N.~{Enrique
  Yalta Soplin}, J.~Heymann, M.~Wiesner, N.~Chen, A.~Renduchintala, and
  T.~Ochiai, ``Espnet: End-to-end speech processing toolkit,'' in \emph{Proc.
  of Interspeech}, 2018, pp. 2207--2211.

\bibitem{zeiler2012adadelta}
M.~D. Zeiler, ``Adadelta: an adaptive learning rate method,'' \emph{arXiv
  preprint arXiv:1212.5701}, 2012.

\end{thebibliography}


\begin{thebibliography}{}
\providecommand{\url}[1]{#1}
\csname url@samestyle\endcsname
\providecommand{\newblock}{\relax}
\providecommand{\bibinfo}[2]{#2}
\providecommand{\BIBentrySTDinterwordspacing}{\spaceskip=0pt\relax}
\providecommand{\BIBentryALTinterwordstretchfactor}{4}
\providecommand{\BIBentryALTinterwordspacing}{\spaceskip=\fontdimen2\font plus
\BIBentryALTinterwordstretchfactor\fontdimen3\font minus
  \fontdimen4\font\relax}
\providecommand{\BIBforeignlanguage}[2]{{%
\expandafter\ifx\csname l@#1\endcsname\relax
\typeout{** WARNING: IEEEtran.bst: No hyphenation pattern has been}%
\typeout{** loaded for the language `#1'. Using the pattern for}%
\typeout{** the default language instead.}%
\else
\language=\csname l@#1\endcsname
\fi
#2}}
\providecommand{\BIBdecl}{\relax}
\BIBdecl

\end{thebibliography}

\end{document}